\documentclass[12pt]{article}

\title{Quantum Parrondo's Games}
\author{A.P. Flitney
\thanks{aflitney@physics.adelaide.edu.au},
J. Ng
\thanks{jng@physics.uq.edu.au}
and D. Abbott
\thanks{dabbott@eleceng.adelaide.edu.au} \\
${}^{* \dag \ddag}$ {\small \em Centre for Biomedical Engineering (CBME)} \\
{\small \em and Dept.\ of Electrical and Electronic Engineering,} \\
{\small \em Adelaide University, SA 5005, Australia} \\
${}^{\dag}$ {\small \em Centre for Quantum Computer Technology,
Dept.\ of Physics,} \\
{\small \em University of Queensland, St Lucia, Qld 4072, Australia}}
\date{}

\begin{document}

\maketitle

\begin{abstract}
Parrondo's Paradox arises when two losing games are combined
to produce a winning one.
A history dependent quantum Parrondo game is studied
where the rotation operators that represent the toss of a classical biased coin
are replaced by general $SU(2)$ operators to transform the game into
the quantum domain.
In the initial state, a superposition of qubits can be used to couple the games
and produce interference leading to 
quite different payoffs to those in the classical case.
\end{abstract}

\noindent
pacs: 03.67.-a, 02.50.le \\
keywords: quantum games, Parrondo's paradox

\section{Introduction}
Game theory is the study of the competing strategies of agents 
involved in some interaction.
First introduced by von Neumann~\cite{neumann51},
it is now widely used in fields as diverse as economics and biology.
Recently, interest has been focused on recasting classical game
theory to the quantum realm in order to study the problems
of quantum information, communication and computation.
The problem of creating useful algorithms for quantum computers
is a difficult one and the study of quantum games
may provide some useful insight.
Meyer~\cite{meyer99} performed the original work in this field in 1999
and since then a number of authors have tackled
coin tossing games~\cite{meyer99,ng01},
the Prisoners Dilemma~\cite{eisert99,benjamin00a,li01a,du01,iqbal01a},
the Battle of the Sexes~\cite{marinatto00,du00a},
the Monty Hall game~\cite{li01b,flitney01},
Rock-Scissors-Paper~\cite{iqbal01b}
and others~\cite{du00b,benjamin00b,kay01,iqbal01c,johnson01}.
Effects not seen in classical game theory can arise
as a result of quantum interference and quantum entanglement.

\section{Parrondo's paradox}
A Parrondo's game is an apparent paradox in game theory
where two games that are losing when played individually
can be combined to produce a winning game.
The effect is named after its discoverer,
Juan Parrondo~\cite{harmer99a,mcclintock99},
and can be mimiced in a physical system of a
Brownian ratchet and pawl~\cite{feynman63,harmer01}
which is apparently driven in one direction
by the Brownian motion of surrounding particles.
The classical Parrondo game is cast in the form of a gambling game
utilising a set of biased coins~\cite{harmer01,harmer99b,parrondo00}.
In this, game $A$ is the toss of a single biased coin
while game $B$ utilises two or more biased coins
whose use depends on the game situation.
The paradox requires a form of feedback,
for example through the dependence on
capital~\cite{harmer99b},
through history dependent rules~\cite{parrondo00},
or through spatial neighbour dependence~\cite{toral01}.
In this paper game $B$ is a history dependent game utilising four coins
$B_1$ to $B_4$ as indicated in Fig.~\ref{fig-games}.

\section{A quantum Parrondo game}
Meyer and Blumer~\cite{meyer01} use a quantum lattice gas to consider
a Parrondo's game in the quantum sphere.
However, consistent with the original idea of Meyer~\cite{meyer99},
and following Ng~\cite{ng01},
we shall quantise the coin tossing game directly
by replacing the rotation of a bit,
representing a toss of a classical coin,
by an $SU(2)$ operation on a qubit.
A physical interpretation of our system could be
a collection of
polarised photons
where $|0\rangle$ represents horizontal polarisation
and $|1\rangle$ represents vertical polarisation
(though we could just as easily consider instead the spin of a spin one-half
particle).

In classical gambling games there is a random element,
and in a Parrondo's game the results of the random process
is used to alter the evolution of the game.
The quantum mechanical model is deterministic until
a measurement is made at the end of the process.
The element of chance that is necessary in the classical game
is replaced by a superposition
that represents all the possible results in parallel.
We can get new behaviour by the addition of phase factors in our
operators and by interference between states.
A further random element can be introduced,
in future studies,
by perturbing the system with noise~\cite{johnson01}
or by considering decoherence during the evolution of the sequence of games.

An arbitrary $SU(2)$ operation on a qubit can be written as
\begin{eqnarray}
\label{eqn-A}
\hat{A}(\theta, \gamma, \delta) &=& \hat{P}(\gamma) \, \hat{R} (\theta)
	\, \hat{P}(\delta) \\ \nonumber
 &=& \left( \begin{array}{cc}
            e^{-i (\gamma + \delta)/2} \cos \theta & -e^{-i (\gamma -
\delta)/2}
		\sin \theta \\
            e^{i (\gamma - \delta)/2} \sin \theta & e^{i (\gamma + \delta)/2}
		\cos \theta \\
     \end{array} \right) \;,
\end{eqnarray}
where $\theta \in [-\pi, \pi]$ and $\gamma, \delta \in [0, 2 \pi]$.
This is our game $A$:
the quantum analogue of a single toss of a biased coin.
One way of achieving this physically on a polarised photon
would be to sandwich a rotation of the plane of polarisation by $\theta$ (R)
between two birefringent media (P) that introduce phase differences of $\gamma$
and
$\delta$, respectively, between the horizontal and vertical planes of
polarisation.
Game $B$ consists of four $SU(2)$ operations,
each of the form of Eq.~\ref{eqn-A},
whose use is controlled by the results of the previous two games
(see Fig.~\ref{fig-games}):
\begin{eqnarray}
\lefteqn{\hat{B}
        (\phi_1, \alpha_1, \beta_1, \phi_2, \alpha_2, \beta_2,
		\phi_3, \alpha_3, \beta_3, \phi_4, \alpha_4, \beta_4) =}\\
\nonumber
 && \makebox[-2mm]{} \left[ \begin{array}{cccc}
        A(\phi_1, \alpha_1, \beta_1) & 0 & 0 & 0 \\
        0 & A(\phi_2, \alpha_2, \beta_2) & 0 & 0 \\
        0 & 0 & A(\phi_3, \alpha_3, \beta_3)  & 0 \\
        0 & 0 & 0 & A(\phi_4, \alpha_4, \beta_4)
     \end{array} \right] \;.
\end{eqnarray}
This acts on the state
\begin{equation}
|\psi(t-2) \rangle \otimes |\psi(t-1) \rangle \otimes |i \rangle \;,
\end{equation}
where $|\psi(t-1) \rangle $ and $|\psi(t-2) \rangle$
represent the results of the two previous games
and $|i \rangle$ is the initial state of the target qubit.
That is,
\begin{equation}
\hat{B} |q_1 q_2 q_3 \rangle = |q_1 q_2 b \rangle \;,
\end{equation}
where $q_1, q_2, q_3 \in \{0,1\}$ and $b$ is the output of the game $B$.

The results of $n$ successive games of $B$ can be computed by
\begin{eqnarray}
|\psi_f \rangle &=&
        (\hat{I}^{\otimes n-1} \otimes \hat{B})
        (\hat{I}^{\otimes n-2} \otimes \hat{B} \otimes \hat{I})
        (\hat{I}^{\otimes n-3} \otimes \hat{B} \otimes \hat{I}^{\otimes 2}) \\
\nonumber
 && \dots
        (\hat{I} \otimes \hat{B} \otimes \hat{I}^{\otimes n-2})
        (\hat{B} \otimes \hat{I}^{\otimes n-1})
        \, |\psi_i \rangle \;,
\end{eqnarray}
with $| \psi_i \rangle$ being an initial state of $n+2$ qubits.
The first two qubits of $| \psi_i \rangle$ are left unchanged
and are only necessary as an input to the first game of $B$.
In this and Eq.~(\ref{eqn-AAB}),
$\hat{I}$ is the identity operator for a single qubit.
The flow of information in this protocol is shown in Fig.~\ref{fig-info}(a).
The result of other game sequences can be computed in a similar manner.
The simplist case to study is that of
two games of $A$ followed by one game of $B$,
since the results of one set of games do not feed into the next.
The sequence $AAB$ played $n$ times results in the state
\begin{eqnarray}
\label{eqn-AAB}
|\psi_f \rangle &=&
	\left( \hat{I}^{\otimes 3n-3} \otimes
	(\hat{B} (\hat{A} \otimes \hat{A} \otimes \hat{I})) \right) \\
\nonumber
 && \makebox[5mm]{} \left(\hat{I}^{\otimes 3n-6} \otimes
		( \hat{B} (\hat{A} \otimes \hat{A} \otimes \hat{I}))
		\otimes \hat{I}^{\otimes 3} \right) \\ \nonumber
 && \makebox[5mm]{} \ldots \left( (\hat{B}
		(\hat{A} \otimes \hat{A} \otimes \hat{I})) \otimes
		\hat{I}^{\otimes 3n-3} \right) \, |\psi_i \rangle \\ \nonumber
&=& \hat{G}^{\otimes n} |\psi_i \rangle \;,
\end{eqnarray}
where $\hat{G} = \hat{B} (\hat{A} \otimes \hat{A} \otimes \hat{I})$
and $| \psi_i \rangle$ is an initial state of $3 n$ qubits.
The information flow for this sequence is shown in Fig.~\ref{fig-info}(c).

In quantum game theory the standard protocol is to take the initial state
$|00 \dots 0 \rangle$,
apply an entangling gate, then the operators associated with the players
strategies and finally a dis-entangling gate~\cite{eisert99}.
A measurement on the resulting state is taken
and then the payoff is determined.
If the entangling gate depends upon some parameter,
then the classical game can be reproduced when this
parameter is set to zero,
representing no entanglement.
In the present case this is problematic
since the entangling gate $\hat{J}$
used by Eisert~\cite{eisert99} and
others~\cite{li01a,du01,benjamin00b,kay01,johnson01}
does not commute with the classical limit
(all phases $\rightarrow 0$)
of $\hat{B}$, which was Eisert's motivation for the choice of $\hat{J}$.
Thus this protocol would not reproduce the classical game when the phases
are set to zero.
So instead we follow~\cite{marinatto00}
and suppose the initial state is already in the maximally entangled state:
\begin{equation}
|\psi_i^{m} \rangle =
	\frac{1}{\sqrt{2}} ( \, |00 \ldots 0 \rangle \,+\,
		|11 \ldots 1 \rangle \, ) \;.
\end{equation}
The classical game can be reproduced by choosing the alternative initial state
$|\psi_i\rangle = |00 \ldots 0\rangle$.
Thus the classical game is still a subset of the quantum one.
If $| \psi_i \rangle$ is a superposition,
interference effects
that either enhance or reduce the success of the player
can be obtained.
The addition of non-zero phases in the
operators $\hat{A}$ and $\hat{B}$ 
can modify this interference.

To determine the payoff let the payoff for a $|1 \rangle$ state be one,
and for a $|0 \rangle$ state be negative one.
The expectation value of the payoff from a sequence of games
resulting in the state $|\psi_f \rangle$
can be computed by
\begin{equation}
\langle \$ \rangle = \sum_{j=0}^{n} \left( (2 j - n)
	\sum_{j'} \left| \langle \psi_{j}^{j'} | \psi_f \rangle
\right|^2\right) \;,
\end{equation}
where the second summation is taken over all basis states
$\langle \psi_j^{j'} | $ with $j$ $1$'s
and $n-j$ 0's.

\section{Results}
Consider the game sequence $AAB$.
With an initial state of $|000 \rangle$
this yields a payoff of
\begin{eqnarray}
\langle \$_{AAB}^{0} \rangle &=&
	\sin^4 \theta \, (2 - \cos 2 \phi_4)
		\:-\: \cos^4 \theta \, (2 + \cos 2 \phi_1) \\ \nonumber
 && \makebox[5mm]{} - \frac{1}{4} \sin^2 2 \theta \,
	(\cos 2 \phi_2 + \cos 2 \phi_3) \;,
\end{eqnarray}
which is the same as the classical result.
In order to get interference there needs to be two different ways
of arriving at the same state.
We need only choose some superposition
not the maximally entangled state,
however this is the most interesting initial state to study.
Choosing $|\psi_i^m \rangle = \frac{1}{\sqrt{2}} (|000 \rangle + |111 \rangle)$
the result is
\begin{eqnarray}
\label{eqn-AABpayoff}
\langle \$_{AAB}^{m} \rangle &=&
\frac{1}{2} \cos 2 \theta \: (\cos 2 \phi_4 - \cos 2 \phi_1) \\ \nonumber
 && \makebox[5mm]{} + \frac{1}{4} \sin^2 2 \theta \,
	\left( \cos(2 \delta + \beta_1)
	\, \sin 2 \phi_1 \right. \\ \nonumber
 && \makebox[10mm]{} - \cos(2 \delta + \beta_2) \, \sin 2 \phi_2
			 \:-\: \cos(2 \delta + \beta_3) \, \sin 2 \phi_3 \\
\nonumber
 && \makebox[10mm]{} \left. + \cos(2 \delta + \beta_4) \, \sin 2 \phi_4 \right)
\;.
\end{eqnarray}
It is the dependence on the phase angles $\delta$ and $\beta_i$
that can produce a result that cannot be obtained in the classical game.
In the quantum case a range of payoffs can be obtained for a given
set of $\theta$ and $\phi_i$'s,
that is, for a given set of probabilities for games $A$ and $B$.

The probabilities given in Fig.~\ref{fig-games} yield
a situation where both games $A$ and $B$ are individually
losing but the combination of $A$ and $B$ can produce a net positive
payoff provided $\epsilon < 1/168$~\cite{parrondo00}.
With the quantum version of the games the expectation value of the payoff
(to $O[\epsilon]$)
for a single sequence of $AAB$ can vary between
$0.812 + 0.24 \epsilon$ and $-0.812 + 0.03 \epsilon$.
The maximum result is obtained by setting
$\beta_2 = \beta_3 = \pi - 2 \delta$ and $\beta_1 = \beta_4 = -2 \delta$,
while the minimum is obtained by
$\beta_1 = \beta_4 = \pi - 2 \delta$ and $\beta_2 = \beta_3 = -2 \delta$.
The values of the $\alpha_i$'s are not relevant.
Classically $AAB$ is a winning sequence provided
$\epsilon < 1/112$ (see Table~\ref{tab-results}).

The average payoff for the classical game sequence $AAB_1$
(that is, $AAB$ where each branch of $B$ is the best branch $B_1$)
is $4/5 - 6 \epsilon$
which is less than the greatest value of $\langle \$_{AAB}^{m} \rangle$.
Thus the entanglement and the resulting interference
can make game $B$ in the sequence $AAB$ better
than its best branch taken alone.
Indeed the expectation value for the payoff of
a quantum $AAB_1$ on the maximally entangled initial state
vanishes due to destructive interference.
(This can be seen from Eq.~(\ref{eqn-AABpayoff})
by setting all the $\phi_i$'s equal to $\phi_1$
and all the $\beta_i$'s to $\beta_1$.)

The quantum enhancement disappears when we play a sequence of $AAB$'s
on the maximally entangled initial state.
In this case the phase dependent terms undergo destructive interference
and we are left with a gain per qubit of order $\epsilon$ (see
Table~\ref{tab-results}).

A sequence of $B$'s leaves the first two qubits unaltered
while a sequence of $AB$'s leaves the first qubit unaffected.
In these cases the final states that arise from
$|\psi_i\rangle = |000\rangle$ and $|\psi_i\rangle = |111\rangle$
are distinct so a superposition of these two states produces no interference.
An initial state that is a different superposition may give interference
effects.

\section{Conclusion}
We have developed a protocol for
a quantum version of a history dependent Parrondo's game.
If the initial state is a superposition,
payoffs different from the classical game can be obtained
as a result of interference.
In some cases payoffs can be considerably altered by adjusting the phase
factors associated with the operators
without altering the amplitudes
(and hence the associated classical probabilities).
If the initial state is simply $|00 \dots 0\rangle$
the payoffs are independent
of the phases and are no different from the classical ones
(with an initial history of loss, loss).
In other cases we may obtain much larger or smaller payoffs
provided the initial state involves a superposition that gives the
possibility of interference for that particular game sequence.

Neil Johnson of Oxford University is gratefully acknowledged for 
pointing out errors in the earlier versions of our manuscript.
This work was supported by GTECH Corporation Australia
with the assistance of the SA Lotteries Commission (Australia).

\pagebreak
\pagestyle{empty}
\begin{figure}
\begin{picture}(400,300)(30,0)
\put(150,175){
\begin{picture}(100,100)(0,0)
	\put(30,90){\bf game A}
	\put(35,50){\line(1,2){15}}
	\put(50,80){\line(1,-2){15}}
	\put(10,60){$\frac{1}{2} + \epsilon$}
	\put(70,60){$\frac{1}{2} - \epsilon$}
	\put(20,35){lose}
	\put(70,35){win}
\end{picture}}
\put(50,135){
\begin{picture}(300,40)(0,-5)
	\put(0,0){\line(1,0){300}}
	\multiput(0,0)(100,0){4}{\line(0,-1){10}}
	\put(100,10){previous two results}
	\put(120,30){\bf game B}
\end{picture}}
\put(0,0){
\begin{picture}(100,120)(0,0)
	\put(23,110){lost, lost}
	\put(50,105){\line(0,-1){10}}
	\put(45,85){$B_1$}
	\put(35,50){\line(1,2){15}}
	\put(50,80){\line(1,-2){15}}
	\put(8,60){$\frac{1}{10} + \epsilon$}
	\put(68,60){$\frac{9}{10} - \epsilon$}
	\put(20,35){lose}
	\put(70,35){win}
\end{picture}}
\put(100,0){
\begin{picture}(100,120)(0,0)
	\put(23,110){lost, won}
	\put(50,105){\line(0,-1){10}}
	\put(45,85){$B_2$}
	\put(35,50){\line(1,2){15}}
	\put(50,80){\line(1,-2){15}}
	\put(10,60){$\frac{3}{4} + \epsilon$}
	\put(68,60){$\frac{1}{4} - \epsilon$}
	\put(20,35){lose}
	\put(70,35){win}
\end{picture}}
\put(200,0){
\begin{picture}(100,120)(0,0)
	\put(25,110){won, lost}
	\put(50,105){\line(0,-1){10}}
	\put(45,85){$B_3$}
	\put(35,50){\line(1,2){15}}
	\put(50,80){\line(1,-2){15}}
	\put(10,60){$\frac{3}{4} + \epsilon$}
	\put(68,60){$\frac{1}{4} - \epsilon$}
	\put(20,35){lose}
	\put(70,35){win}
\end{picture}}
\put(300,0){
\begin{picture}(100,120)(0,0)
	\put(25,110){won, won}
	\put(50,105){\line(0,-1){10}}
	\put(45,85){$B_4$}
	\put(35,50){\line(1,2){15}}
	\put(50,80){\line(1,-2){15}}
	\put(8,60){$\frac{3}{10} + \epsilon$}
	\put(68,60){$\frac{7}{10} - \epsilon$}
	\put(20,35){lose}
	\put(70,35){win}
\end{picture}}
\end{picture}
\caption{}
\label{fig-games}
\end{figure}

\pagebreak
\begin{figure}
\setlength{\unitlength}{1.5pt}
\begin{picture}(200,360)(-25,0)
\put(0,240){
\begin{picture}(185,120)(-35,-10)

\put(-15,34){\framebox(25,71){$| \psi_i \rangle$}}
\multiput(10,103)(0,-8){2}{\line(1,0){140}}
\put(10,87){\line(1,0){15}}
\multiput(25,82)(25,-8){3}{\framebox(10,10){$\hat{B}$}}
\put(35,87){\line(1,0){115}}
\put(10,79){\line(1,0){40}}
\put(60,79){\line(1,0){90}}
\put(10,71){\line(1,0){65}}
\put(85,71){\line(1,0){15}}
\put(100,61){\makebox(10,10){$\ddots$}}
\multiput(10,54)(0,-8){2}{\line(1,0){15}}
\multiput(110,54)(0,-8){2}{\line(1,0){40}}
\put(125,33){\framebox(10,10){$\hat{B}$}}
\put(135,38){\line(1,0){15}}
\put(150,34){\framebox(25,71){$| \psi_f \rangle$}}
\multiput(30,103)(25,-8){3}{\circle*{3}}
\multiput(30,95)(25,-8){3}{\circle*{3}}
\multiput(30,103)(25,-8){3}{\line(0,-1){11}}
\multiput(130,54)(0,-8){2}{\circle*{3}}
\put(130,54){\line(0,-1){11}}
\put(-40,70){(a)}

\end{picture}
}


\put(0,120){
\begin{picture}(170,120)(-35,-10)

\put(-15,34){\framebox(25,71){$| \psi_i \rangle$}}
\put(10,103){\line(1,0){150}}
\multiput(10,95)(20,0){2}{\line(1,0){10}}
\put(30,95){\line(1,0){130}}
\put(10,87){\line(1,0){30}}
\put(50,87){\line(1,0){110}}
\multiput(20,90)(40,-16){2}{\framebox(10,10){$\hat{A}$}}
\multiput(40,82)(40,-16){2}{\framebox(10,10){$\hat{B}$}}
\put(10,79){\line(1,0){50}}
\put(70,79){\line(1,0){90}}
\put(10,71){\line(1,0){70}}
\put(90,71){\line(1,0){10}}
\put(100,59){\makebox(10,10){$\ddots$}}
\put(10,52){\line(1,0){15}}
\put(110,52){\line(1,0){50}}
\put(10,44){\line(1,0){110}}
\put(120,39){\framebox(10,10){$\hat{A}$}}
\put(130,44){\line(1,0){30}}
\put(10,36){\line(1,0){130}}
\put(150,36){\line(1,0){10}}
\put(140,31){\framebox(10,10){$\hat{B}$}}
\put(160,34){\framebox(25,71){$| \psi_f \rangle$}}
\multiput(45,103)(0,-8){2}{\circle*{3}}
\multiput(85,87)(0,-8){2}{\circle*{3}}
\multiput(45,103)(40,-16){2}{\line(0,-1){11}}
\multiput(145,52)(0,-8){2}{\circle*{3}}
\put(145,52){\line(0,-1){11}}
\put(-40,70){(b)}

\end{picture}
}

\put(0,0){
\begin{picture}(170,120)(-35,-10)

\put(-15,34){\framebox(25,71){$| \psi_i \rangle$}}
\put(10,100){\line(1,0){10}}
\put(30,100){\line(1,0){120}}
\put(10,92){\line(1,0){30}}
\put(50,92){\line(1,0){100}}
\put(10,84){\line(1,0){50}}
\put(70,84){\line(1,0){80}}
\multiput(20,95)(20,-8){2}{\framebox(10,10){$\hat{A}$}}
\multiput(60,79)(60,-25){2}{\framebox(10,10){$\hat{B}$}}
\put(10,75){\line(1,0){70}}
\put(90,75){\line(1,0){60}}
\put(10,67){\line(1,0){90}}
\put(110,67){\line(1,0){40}}
\put(10,59){\line(1,0){110}}
\put(130,59){\line(1,0){20}}
\multiput(80,70)(20,-8){2}{\framebox(10,10){$\hat{A}$}}
\put(125,40){\makebox(10,10){$\vdots$}}
\put(150,34){\framebox(25,71){$| \psi_f \rangle$}}
\multiput(65,100)(0,-8){2}{\circle*{3}}
\multiput(125,75)(0,-8){2}{\circle*{3}}
\multiput(65,100)(60,-24){2}{\line(0,-1){11}}

\put(30,10){\vector(1,0){100}}
\put(70,0){time}
\put(-40,70){(c)}

\end{picture}
}
\end{picture}
\caption{}
\label{fig-info}
\end{figure}
\clearpage

\begin{table}
\begin{tabular}{|l|c|cc|}
\hline
sequence & classical payoff & quantum payoff & \\
\hline
AA \dots A & $-2 \epsilon$ & 0 & \\
B & $1/60 - 2 \epsilon/3$ & $1/15$ & \\
BB & $1/75 - 19 \epsilon/15$ & $13/400 + \epsilon/20$ & \\
BBB & $0.008 - 1.1 \epsilon$ & $0.017 + 0.03 \epsilon$ & \\
AB & $1/60 - 19 \epsilon/15$ & $1/30 + \epsilon/15$ & \\
ABAB & $0.032 - 2.5 \epsilon$ & $0.019 + 0.08 \epsilon$ & \\
AAB & $1/60 - 28 \epsilon/15$ & $-0.271 + 0.03 \epsilon \,;$ &
	$0.271 + 0.24 \epsilon$ \\
AAB \dots AAB & $1/60 - 28 \epsilon/15$ & $2 \epsilon/15$ & \\
\hline
\end{tabular}
\caption{}
\label{tab-results}
\end{table}
\clearpage

\noindent
Figure captions:

\begin{enumerate}
\item Winning and losing probabilities for game $A$
and the history dependent game $B$ from
Parrondo, Harmer and Abbott~\cite{parrondo00}.
\item
The information flow in qubits (solid lines) in a sequence of
(a) $B$,
(b) an alternating sequence of $A$ and $B$,
and (c) two games of $A$ followed by one of $B$.
Note in (c) that the output of one set of $AAB$ does not feed into the next.
In each case a measurement on $|\psi_f \rangle$ is taken on completion of the
sequence of games to determine the payoff.
\end{enumerate}
Table captions:
\begin{enumerate}
\item Expectation values for the payoff
per qubit to $O[\epsilon]$ for various sequences of games.
The classical payoffs are the average over the possible initial conditions
(that is, the results of the two previous games for sequences of $B$
and the results of the previous game for sequences of $AB$),
while the quantum payoffs are calculated for the
maximally entangled initial state,
$\frac{1}{\sqrt{2}} (|00 \ldots 0 \rangle + |11 \ldots 1 \rangle)$.
For the sequence AAB the two values given for the quantum payoff
are the minimum and maximum, respectively (see text).
\end{enumerate}

\end{document}